# Radiative-channel valley topological laser

Seonyeong Kim[1,2,3,4], Markus Scherrer[4], Jakub Dranczewski[4,5], Heinz Schmid[4], Kirsten Moselund[2,3,*], and Chang-Won Lee[1,*]

[1]Institute of Advanced Optics and Photonics, Hanbat National University, Daejeon, 34158, Korea

[2]Paul Scherrer Institute, Forschungsstrasse 111, Villigen 5232, Switzerland

[3]École Polytechnique Fédérale de Lausanne (EPFL), STI INPhO, Lausanne, Switzerland

[4]IBM Research Europe – Zürich, Säumerstrasse 4, Rüschlikon 8803, Switzerland

[5]Blackett Laboratory, Department of Physics, Imperial College London, SW7 2BW London, UK

*Corresponding authors: kirsten.moselund@psi.ch and cwlee42@hanbat.ac.kr

**Abstract**

Active topological photonic systems have opened a new landscape for robust light control, offering new pathways for semiconductor lasing enabled by topology-driven effects. However, their inherently non-Hermitian nature—combining gain, radiation leakage, and material loss—makes the underlying physics far more complex, and prior studies have mostly focused on gain competition, leaving the influence of loss channels less examined. Here, we experimentally demonstrate radiative-channel-driven topological lasing in a valley photonic crystal consisting of isolated InP nanorods on insulator, achieving room-temperature single-mode operation within a ~4λ-scale cavity. Loss-included simulations show that material absorption and radiative leakage can be effectively exploited for establishing the lasing pathway. Local off-edge pumping confirms spatial evidence of topological edge-guided lasing. Berry-curvature calculations, reflecting the unit-cell symmetry breaking, verify the valley-Hall interface in the triangular lattice. Geometry and temperature tuning identify a narrow spectral window where the above-light-line edge branch matches the gain–loss balance and remains decoupled from the bulk bands. Our results elucidate the decisive role of the loss landscape in governing topological lasing and establish a radiative-edge design framework for active topological photonics, while the simple, substrate-supported nanorod architecture additionally provides a scalable platform for implementing these concepts on chip.

## Introduction

Topological photonics have fundamentally altered our approach to controlling light, enabling the realization of optical states that are immune to backscattering and robust against structural disorder[1]. By translating concepts from condensed matter physics—such as the quantum Hall[2] and quantum spin Hall effects[3]—into the optical domain, researchers have successfully demonstrated topological lasers that combine high coherence with the resilience of topological protection. Topological lasers built on valley photonic crystals (VPCs) exploit valley-locked edge states formed by inversion-symmetry breaking, enabling light to confine robustly along domain walls[4–15]. In particular, VPCs allow topological confinement to emerge from minimal geometric variations in dielectric lattices, leading directly to bandgap opening and compact cavity formation[16]. This structural simplicity, together with the intrinsic robustness of valley-dependent edge channels, makes VPC-based topological lasers particularly attractive for compact photonic integrated circuits[17,18].

Topological lasing intrinsically requires an active medium and therefore operates in a non-Hermitian regime, where optical gain, material absorption, and radiative leakage coexist and interact[19]. Owing to this complexity, prior studies have primarily focused on gain engineering and mode competition, while the role of loss channels has often been addressed only implicitly or within simplified models. In particular, the interplay between material loss and valley-Hall edge modes has remained only partially explored, despite its potential implications for the behavior of active topological systems. Experimentally demonstrated VPC lasers to date mostly rely on hole-type suspended photonic-crystal membranes[7,8,15], which combine in-plane guidance by the PhC lattice and total internal reflection in the vertical direction determined by the refractive index between the membrane and the surrounding air. The high index contrasts readily achievable in common semiconductor platforms make it possible to achieve high Q factors[20]. Optically pumped VPC lasing generally employs either global illumination of the entire structure[7,8,15] or shaped pumping along the edges of the cavity either by the use of a spatial light modulator[13], or using grating couplers to couple light to the edge-channel[4]. The entire lasing cavity in these approaches is pumped at once. In contrast, rod-based platforms inherently suffer from stronger out-of-plane radiation loss compared with air-hole membranes[21], placing topological lasing in a fundamentally open and strongly radiative regime. Rod-type VPCs have been extensively explored at acoustic[22–26], microwave[27–34], and terahertz[35–40] frequencies, whereas in the optical regime only a few theoretical studies have been reported[41–43], and the implications of such an intrinsically radiative environment for topological lasing remain largely unexplored.

Here, we demonstrate a rod-type VPC topological laser operating at room-temperature in an intrinsically open non-Hermitian regime. The design is guided by band-structure and Berry-curvature calculations, which confirm the topological origin of the valley-Hall interface and motivate the choice of a triangular-lattice geometry. By systematically varying the cavity geometry and temperature, we identify the conditions under which lasing selectively emerges from the radiative segment of the topological edge mode, as corroborated by loss-inclusive electromagnetic simulations. Importantly, lasing is not determined by the existence of an edge state alone, but by the alignment of its radiative segment with the gain–loss balance while remaining sufficiently decoupled from bulk bands. Despite the compact cavity size of only ~4λ and the open nature of the vertical rods, the device exhibits low-threshold operation, high slope efficiency, and a side-mode suppression ratio (SMSR) of ~30 dB, highlighting high spectral purity. Local excitation slightly away from the cavity edge produces a lasing mode confined to the topological edges, thereby offering clear evidence of edge-guided lasing beyond trivial localized emission. We benchmark our topological-edge laser against a conventional 2D PhC band-edge laser on the same device, showing clear performance enhancement. These results establish a compact, substrate-supported rod-type valley photonic-crystal platform in which topological confinement and radiative loss jointly govern robust single-mode lasing in an open, non-Hermitian nanophotonic system.

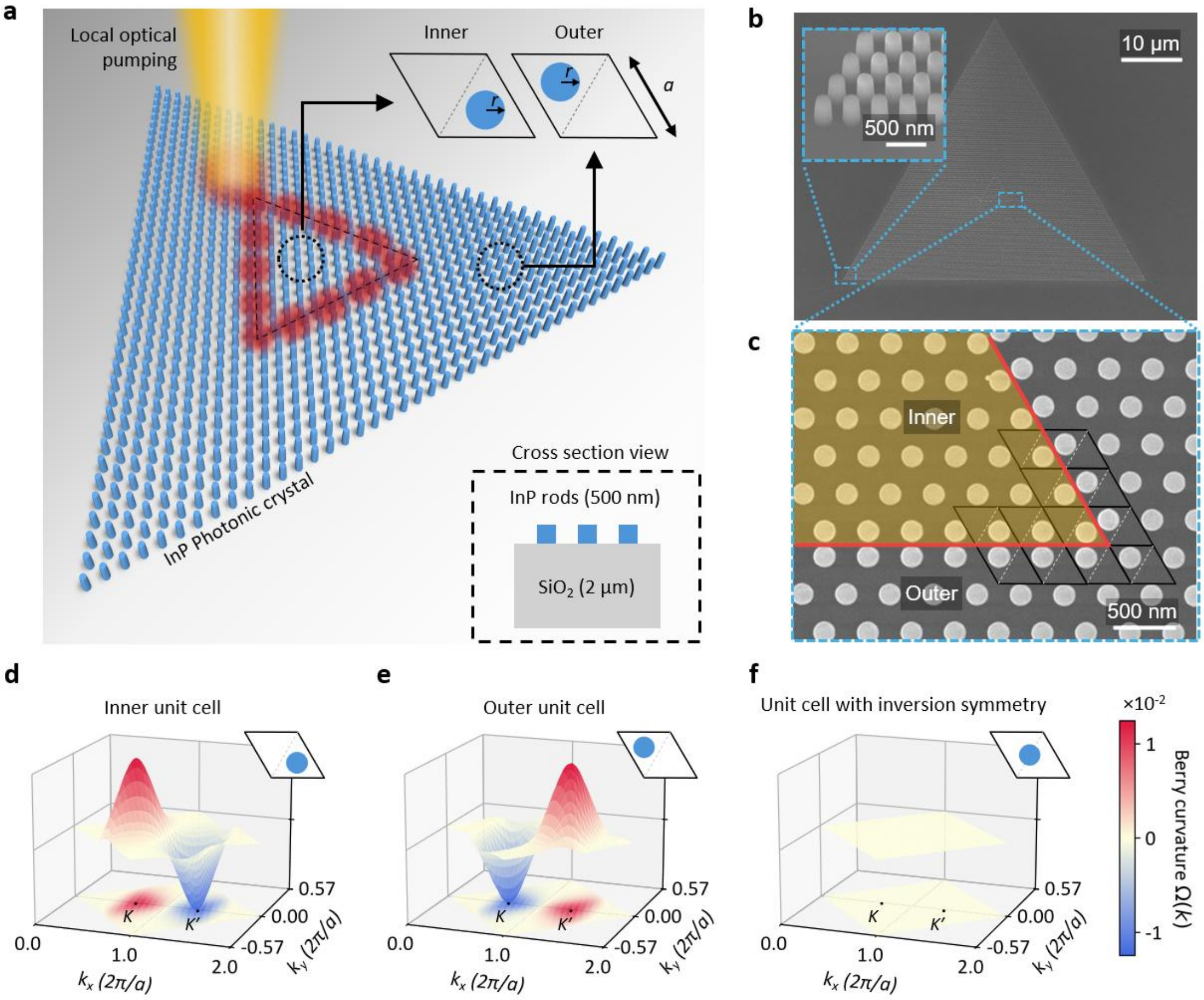

**Fig.1 | Design and schematics of ring-cavity topological laser.**

**a**, Schematic illustration of a triangular ring-cavity topological laser fabricated in a 500-nm-thick InP layer wafer-bonded onto a Si substrate with a 2-μm of $SiO_2$ buffer layer, providing optical isolation. The bottom-right shows the cross-sectional view. The photonic crystal (PhC) consisting of vertically etched InP nanorods supports a photonic band gap in the transverse magnetic (TM) polarization. A rhombic unit cell is defined such that the photonic atom (blue circle) is placed asymmetrically within the unit cell, breaking inversion symmetry. The inner and outer domains of the triangular cavity are formed by inverted unit cell configurations, defining a topological interface at their boundary. Local optical pumping ($\lambda$ = 720 nm, ~3 μm diameter) is focused slightly off the cavity edge, rather than directly on it, to excite interface-confined lasing. **b**, Scanning electron micrograph (SEM) image of the fabricated device. Inset: tilted-view near the lower-left corner showing uniform, well-formed InP nanorods. **c,** Magnified SEM of the right corner highlighted in (b). Away from the interface (red line), both domains are triangular-lattice PhCs with identical parameters ($a$ = 340 nm, $r$ = 0.19$a$ = 64.6 nm), thus sharing the same photonic band. The cavity side length is 4.76 μm (14 unit cells). **d–f**, Calculated Berry-curvature distributions Ω(k) for unit cells with inversion symmetry (f, vanishing curvature) and without inversion symmetry (d,e, strong valley-contrasting curvature).

**Design and excitation geometry of the topological laser.**

**Figure 1a** illustrates the cavity design, highlighting the structural simplicity of a fully substrate-supported valley photonic crystal laser (fabrication details provided in Methods). The SEM image of the fabricated device, shown in **Fig. 1b**, confirms high uniformity and structural fidelity, with well-formed InP nanorods and a clearly visible triangular cavity created by a well-defined photonic crystal (PhC) lattice. The magnified view in **Fig. 1c** captures the cavity corner, illustrating how the distinct configuration of the two-unit cells defines the edge marked by the red line. The chosen unit-cell configuration functionally behaves as a perturbed honeycomb lattice exhibiting the largest photonic band gap[44]. Such a large gap enhances mode confinement and suppresses intervalley scattering, providing favorable conditions for robust topological edge lasing[45]. Considering the bulk regions away from the edge, the basis selection consequently forms a triangular lattice. Previous studies have interpreted the emergence of topological edge states in such a triangular lattice in terms of chiral vortex–valley locking[44] or the symmetry indicator[46], due to their vanishing Berry curvature. In our case, the topological origin of the interface modes is elucidated by Berry-curvature calculations (**Fig. 1d,e**; details in Supplementary **Fig. S1**), which reveal reversed valley-contrasting curvature distributions at K and K′ for the two unit-cell configurations constituting the edge. An inversion-symmetric unit cell yields vanishing curvature (**Fig. 1f**), consistent with earlier reports[11,44,46]. This finding establishes that triangular-lattice PhCs still belong to the valley-Hall, underscoring a general design principle for VPC-based topological edges: the unit cell configuration and domain alignment, rather than global lattice symmetry, determine the dispersion and localization of edge states (See S2 and **Fig. S2** for details).

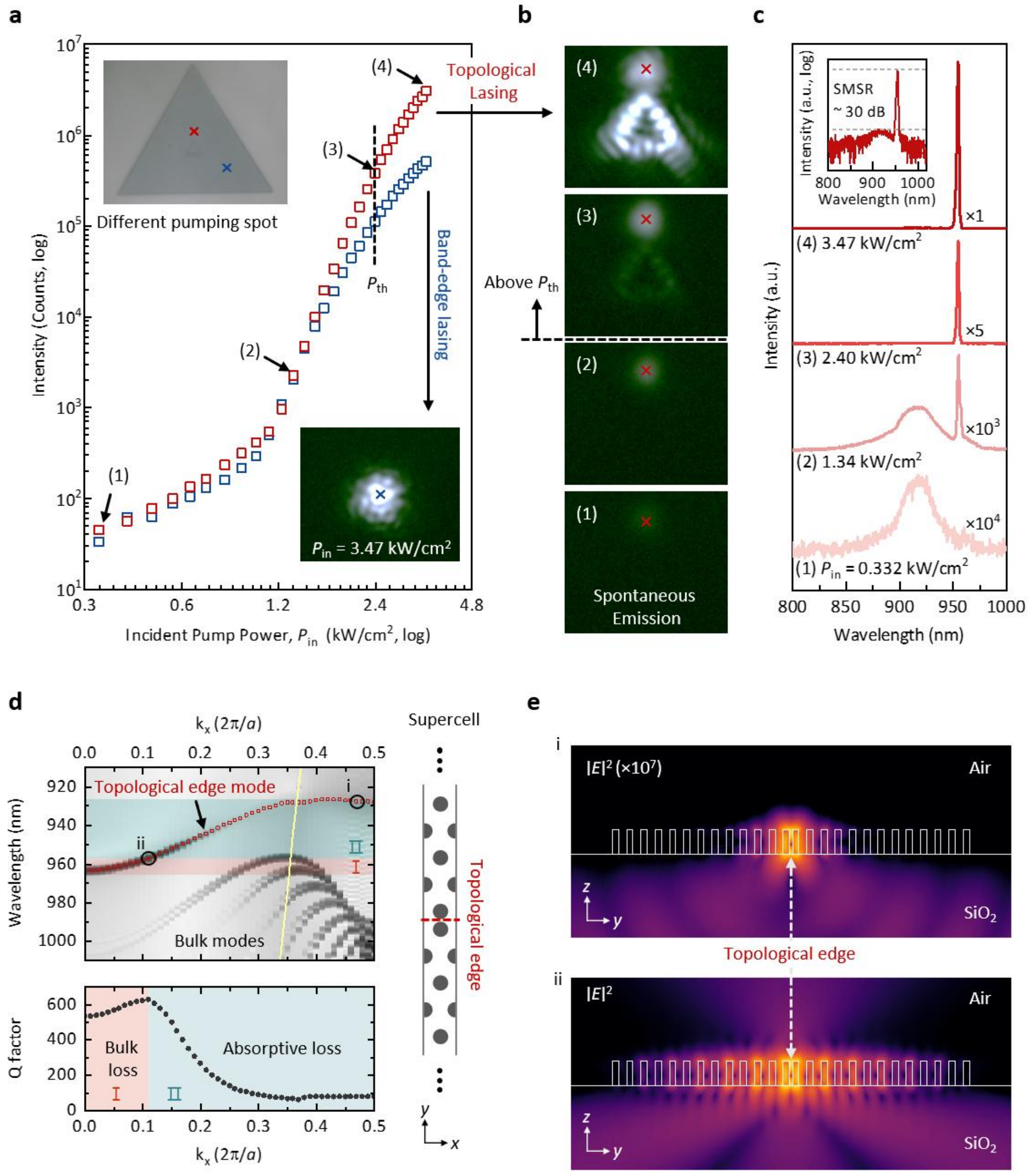

**Fig. 2 | Optical characterization of the fabricated topological laser and comparison with simulation results.**

**a**, Measured output intensity at the lasing peak as a function of incident pump power ($P_{\mathrm{in}}$), obtained by spatially resolved micro-photoluminescence (μ-PL) measurements at room temperature. This pumping geometry enables localized excitation at different positions within the same device. Lattice constant: $a$ = 340 nm; rod radius: $r$ = 0.19$a$. Two distinct lasing modes are observed depending on the pump location (red: topological edge, blue: band-edge), as shown in the optical image in the inset. A similar threshold ($P_{\mathrm{th}}$) is marked by a dashed line for both curves, where $P_{\mathrm{th}}$ is determined from the x-axis intercept of a linear-fit extrapolation in the linear-scale. The emission image of the band-edge laser at saturation is shown in the lower inset. **b**, Power-dependent emission images of the topological laser corresponding to the pump powers in (**a**). The four powers labeled (1)–(4) represent increasing pump powers, from below threshold to the saturation regime. (4) shows clear ring-like emission localized along the triangular-shaped topological edge. **c**, Emission spectra of the topological laser at the corresponding powers. Raw spectra are vertically scaled for clarity: $\times10^4$, $\times10^3$, and ×5 for (1), (2), and (3), respectively. The inset shows the spectrum at maximum power (3.47 kW/cm²) on a logarithmic scale, with a side-mode suppression ratio (SMSR) of approximately 30 dB. **d**, Calculated projected band structure (top) and Q factor (bottom) of the topological edge mode for a supercell geometry with 13 unit cells on each side of the edge (illustrated on the right). The grayscale contour background represents the photonic bands, where darker regions indicate higher mode intensities. The topological edge mode is highlighted by the red markers, and the light line is indicated in yellow. **e**, Cross-sectional electric field intensity profiles $|E|^2$ of the topological edge mode shown in (**d**). (i) Mode with maximum Q below the light line, scaled by $10^7$ for direct comparison. (ii) Mode above the light line.

**Experimental verification of the topological laser.**

The red (blue) curve in **Fig. 2a** was obtained by pumping at the red cross near the cavity edge (blue cross in the uniform PhC region) shown in the inset (Method). Both curves show S-shaped input–output behaviors with a low lasing threshold of ~2.4 $kW/cm^2$. Despite comparable thresholds, the topological laser (red) exhibits a markedly steeper slope efficiency, consistent with more efficient stimulated emission that of the band-edge laser (blue). The lasing dynamics are evident in real-space emission images. For a band-edge laser, the above-threshold emission remains confined to the pump spot (**Fig. 2a**, lower-right inset). For the topological cavity, the CCD sequence in **Fig. 2b** (pump powers (1)–(4) in **Fig. 2a**) shows a transition from diffuse, spontaneous-emission-dominated patterns at low power (1,2) to a bright edge-emission pattern tracing the designed interface above threshold (3,4). Notably, edge lasing is not observed when the excitation spot is placed directly on the cavity edge, indicating that local gain alone is insufficient. Together with the spectral proximity to the band-edge laser, this behavior suggests that off-edge excitation in the periodic bulk PhC region more efficiently populates bulk/band-edge states that feed into the topological edge mode, whereas direct edge pumping does not lead to comparable build-up. The simultaneous spectra in **Fig. 2c** reproduce this spatial transition: the broadband background photoluminescence collapses into a single sharp lasing peak (~955 nm) as the pump crosses threshold. A single dominant peak is already discernible at relatively low excitation, unlike typical waveguide-ring lasers that often require higher pump intensities to reach stable single-mode operation[47]. In the saturation regime (4), the logarithmic-scale spectrum (inset) yields a side-mode suppression ratio (SMSR) of ~30 dB. The band-edge laser shows a similar spectral evolution with a nearly identical peak wavelength, but a lower SMSR (~20 dB; **Fig. S3**). Because the nominal pump power is defined identically in both cases while the excitation pathways differ—direct pumping on the uniform region for the band-edge laser versus indirect, scattering-mediated coupling into the edge mode for the topological cavity—the measured performance of the topological laser should be viewed as a conservative estimate. The detailed performance comparison to other VPC lasers is provided in Table S1.

The calculated projected bands as a function of $k_x$ are shown in **Fig. 2d** (Methods), which shows bulk states and a topological edge branch (red markers) whose visibility progressively fades at shorter wavelengths (i.e., larger $k_x$ in this projection). Here, we used the measured complex refractive index of the InP film ($n + i\kappa$; **Fig. S4**), capturing both vertical radiative leakage and material absorption. The corresponding $Q(k_x)$ profile exhibits two regimes. For $k_x \lesssim 0.11$ (region I), the edge branch overlaps with the bulk bands and partially couples to them, allowing leakage into extended bulk states[48,49]. As $k_x$ increases, this coupling weakens and Q rises to a maximum. At shorter wavelengths (region II), where κ increases sharply as the photon energy approaches the InP band gap, Q falls again in an absorption-limited regime, even below the light line. The measured lasing wavelength lies near a local Q optimum of the edge branch, defining a narrow window where bulk-band hybridization is minimized while absorption is not yet dominant.

**Fig. 2e** compares representative edge modes below and above the light line (marked in **Fig. 2d**). The key difference lies in the field magnitude and the leakage channel. Below the light line, the mode is tightly confined to the interface and well isolated from the bulk branches, but its overall intensity is dramatically reduced by absorption (necessitating a $\times 10^7$ rescaling for visualization). Above the light line, the mode retains edge localization while coupling to free space, providing a vertical radiation channel. Together, these loss-included simulations indicate that lasing is selected from the radiative (above-light-line) segment of the edge dispersion, rather than from a purely non-radiative, below-light-line edge state. This stands in clear contrast to previous topological-lasing studies that treated out-of-plane radiation as detrimental[15,50]. Indeed, when the measured material loss is artificially removed (**Fig. S5**), the edge mode exhibits a moderately higher Q below the light line than above it.

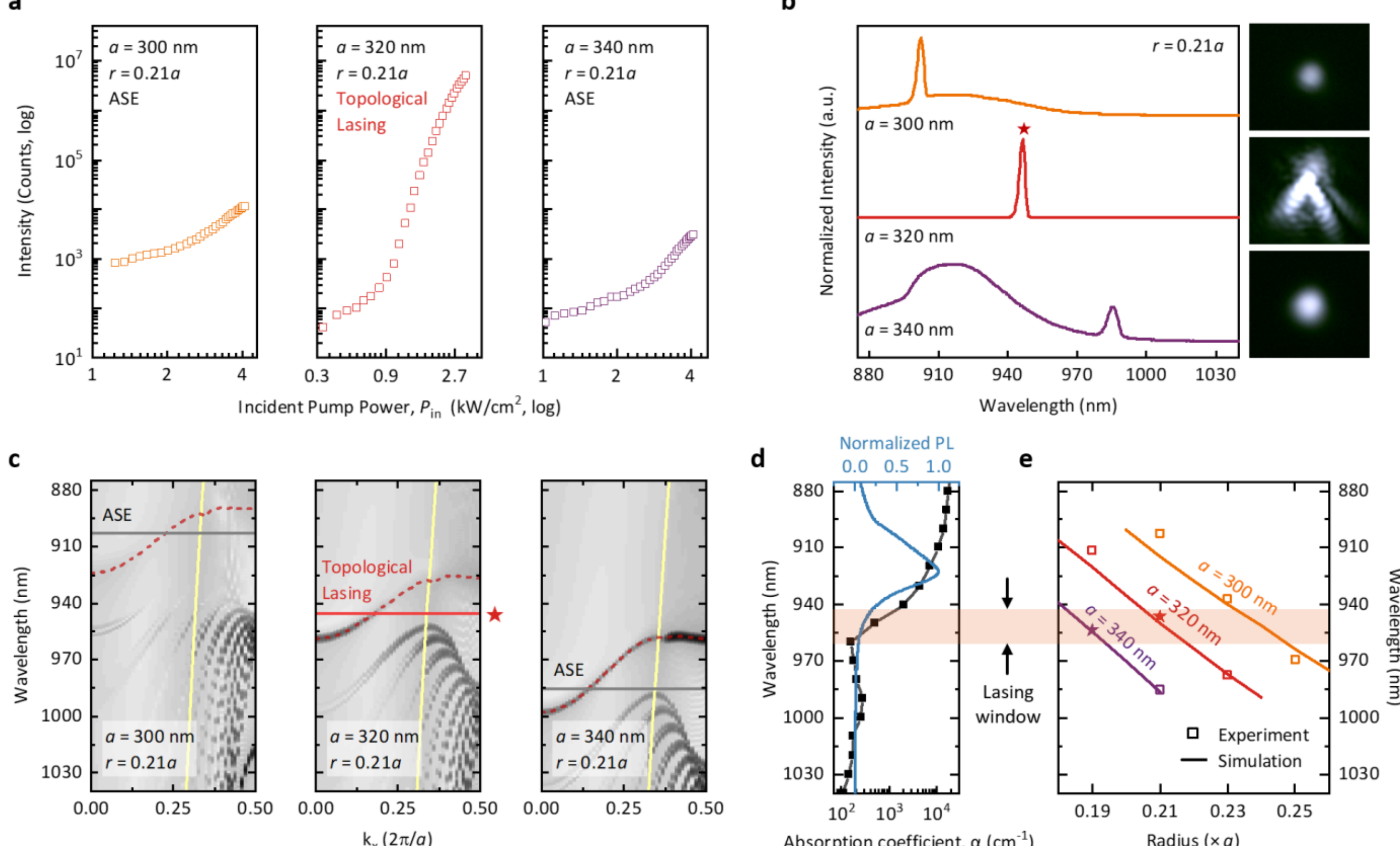

**Fig. 3 | Geometry-dependent emission and the lasing window.**

**a**, Pump–power dependences (log–log) of the intensity at the spectral peak for devices with fixed radius $r = 0.21a$ and lattice constants $a = 300$, 320, and 340 nm. **b**, Emission spectra for the three devices, measured under their respective maximum pump conditions. A red star marks a peak for the 320 nm case, which exhibits an edge-guided emission pattern. Right: corresponding emission patterns acquired simultaneously by a CCD camera. **c**, Corresponding projected band structures. The gray scale represents the mode intensity. The red dashed line serves as a guide to the topological edge mode extracted from the resonance peak center at each $k_x$. The yellow line denotes the light line. **d**. Room-temperature PL (blue) and absorption coefficient α (black) of unpatterned InP layer, serving as reference gain and loss proxy profiles. **e**. Extracted emission peak wavelengths (symbols) for the corresponding devices with different radii and lattice constants (orange for $a = 300$ nm, red for $a = 320$ nm, and purple for $a = 340$ nm). Solid lines indicate the simulated lower band-edge wavelengths for comparison. Stars denote cases where topological lasing is observed.

.

## Geometric Criteria for Topological Lasing

Building on **Fig. 2**, we tune the device geometry to shift the edge dispersion and map a lasing window in parameter space while keeping room temperature. **Fig. 3a** shows the pump–power dependences for devices with fixed normalized radius ($r/a$ = 0.21) but different lattice constants ($a$ = 300, 320, 340 nm). Only the 320 nm device exhibits a clear S-shaped input–output curve and gain clamping, indicating lasing, while the 300 nm and 340 nm devices display only modest superlinear growth, consistent with band-edge amplified spontaneous emission (ASE). **Fig. 3b** presents the corresponding emission spectra and images. All three devices show a distinct resonance that red-shifts with increasing a, as expected from photonic scaling, yet only the 320 nm device produces edge-guided (topological) emission along the designed edge. The propagation does not extend to the entire length of the triangular edge, though, but we speculate this might be related to the non-Hermitian skin effect of the triangular shape[51]. The brighter local emission for $a$ = 340 nm compared to 320 nm, consistent with the higher background PL, is attributed to spontaneous-emission redistribution[52]. This geometric selectivity arises from non-uniform pumping, which makes absorption loss in InP more relevant, as discussed below.

The projected band structures of all three devices are shown in **Fig. 3c**. The measured resonance wavelengths are marked by horizontal lines, while the edge branches and light line are indicated by dashed red and yellow curves, respectively. **Fig. 3d** shows the intrinsic optical response of the unstructured InP film, with room-temperature PL serving as a proxy for available gain and the absorption coefficient α, representing optical loss. For $a$ = 300 nm, both bulk band-edge modes and the topological edge branch lie at short wavelengths and are barely visible owing to strong InP absorption. The red-shifted band dispersion for $a$ = 320 nm closely matches the regime that supports the experimentally observed edge-guided lasing. For $a$ = 340 nm, an instructive feature emerges, as all bands move into a low-absorption region (to wavelengths longer than ~950 nm), the topological modes below the light line evolve into high-Q states, appearing as dark streaks in the map. However, even under these favorable conditions, edge-guided lasing does not originate from these modes. Instead, emission is observed near the photonic band-edge wavelength in the form of ASE. This finding reinforces the mechanism in **Fig. 2** by ruling out edge lasing from below-light-line high-Q modes. In other words, topological edge lasing in our system arises not from maximized in-plane confinement but from radiative coupling through the above-light-line branch.

**Fig. 3e** summarizes the geometric dependence of the resonance wavelength. The measured peaks (symbols) follow the simulated lower band-edge wavelengths (solid lines). Edge-guided topological lasing (stars) appears only within a narrow spectral window (shaded and labeled “lasing window”), whereas devices outside this window exhibit band-edge ASE. Our results indicate that the lasing-eligible region is very much limited to the portion of the edge dispersion above the light line while remaining decoupled from the bulk bands. Geometry tuning then shifts this region relative to the fixed material optical properties, and lasing emerges only when part of this region falls inside the material gain–loss balance window. Under such overlap conditions, the long-wavelength side of the lasing-eligible region experiences the smallest material absorption, so the observed lasing wavelengths naturally appear near the bulk band edge.

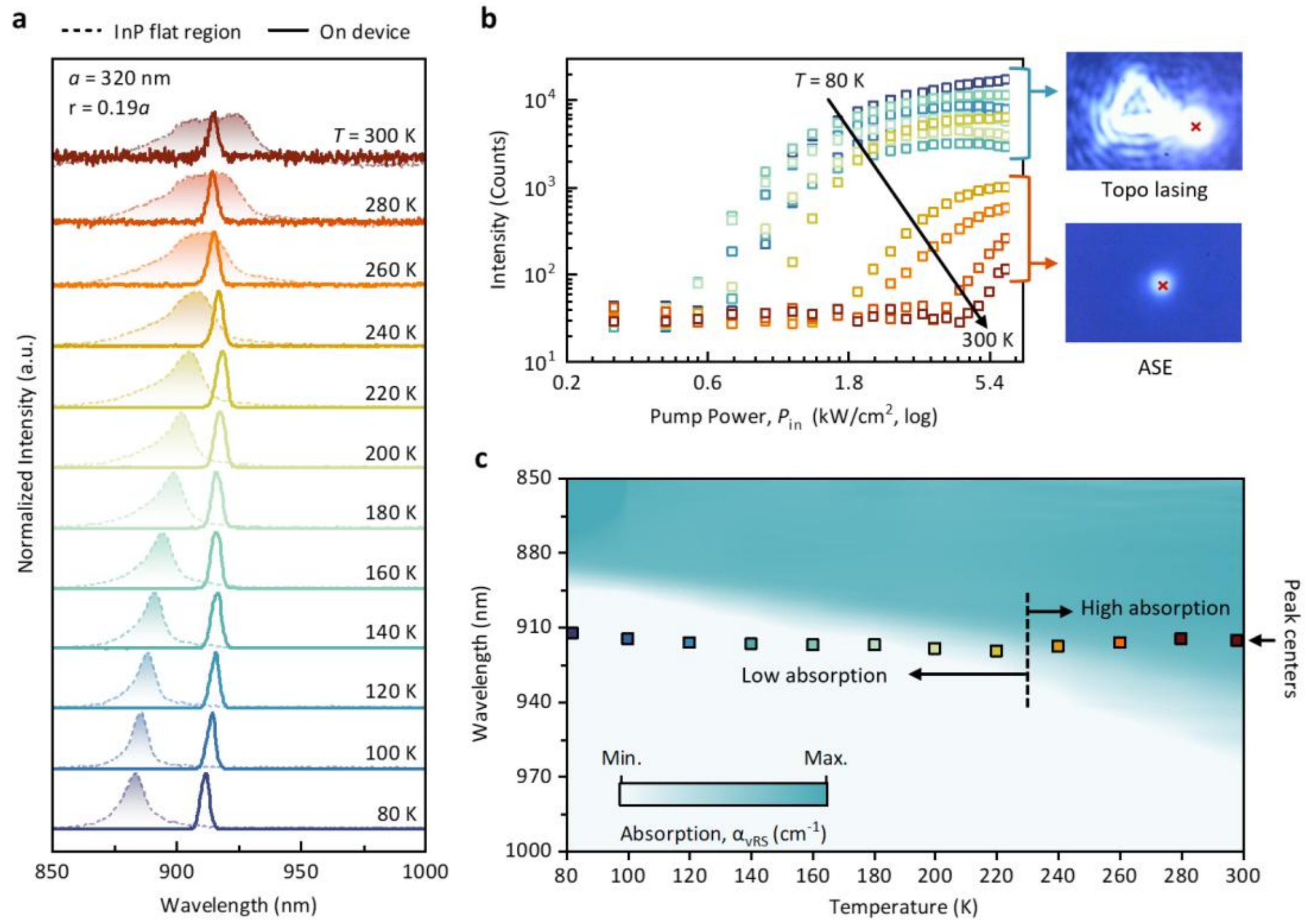

**Fig. 4 | Temperature-driven ASE–lasing crossover and material absorption landscape.**

**a**, Normalized emission spectra measured on the device (solid) for lattice constant $a$ = 320 nm and radius $r$ = 0.19$a$, and on an unpatterned InP flat reference region on the same chip (dashed), over $T$ = 80–300 K in 20 K increments. **b**, L–L (output–pump) characteristics at representative temperatures (colors as in **a**). Right: emission patterns at the same pump density—at high temperatures ($T \approx$ 240–300 K), ASE produces a compact, pump-localized spot, whereas at low temperatures ($T \leq$ 220 K), edge/topological lasing yields a loop-like rim emission along the device perimeter, with the crossover visible near ~240 K. **c**, Temperature–wavelength map with absorption shading $\alpha_{vRS}$ derived from flat-region PL via the van Roosbroeck–Shockley (vRS)-based surrogate serving as a qualitative indicator of relative loss across $T$. Squares mark the emission peak at each temperature.

**Temperature-driven ASE–lasing crossover**

In **Fig. 4a**, we examine temperature-dependent emission spectra (80–300 K) of the device with $a$ = 320 nm and $r$ = 0.19$a$ (solid), which exhibits topological edge lasing only at low temperature, and compare them with spectra from an unpatterned InP reference region on the same chip. The reference PL spectra follow the expected Varshni-type monotonic redshift of the InP band-to-band emission (**Fig. S6**). In contrast, the resonance peak associated with the cavity exhibits only a very weak drift and remains within a narrow wavelength range over the entire temperature span. This disparity in spectral motion highlights that temperature primarily affects the material optical response rather than the cavity resonance. The weak temperature dependence of the cavity resonance is consistent with the fractional change in the InP refractive index over this range ($\Delta n/n \lesssim$ 1–2%)[53,54] and the negligible contribution from thermal expansion ($\Delta L/L \sim 10^{-3}$)[55]. As a result, the temperature-dependent changes observed in the emission behavior mainly reflect variations in the material gain–loss spectrum, while the underlying photonic bands remain largely unchanged.

**Fig. 4b** shows the pump-power dependence at each peak, revealing how strongly the emission behavior diverges with temperature. At high temperatures, the curves display ASE-like behavior without a clear kink, whereas upon cooling a pronounced threshold and steeper post-threshold slope emerge, indicating edge-guided lasing. The change in behavior becomes pronounced near $T$ = 240 K. This contrast is also directly visible in the right-hand optical emission images, where high-temperature devices show only a pump-localized spot, whereas low-temperature devices exhibit a rim-like edge emission.

To further relate this behavior to the material loss landscape, **Fig. 4c** presents a temperature–wavelength map of absorption derived from the flat-region PL using a van Roosbroeck–Shockley (vRS)–based conversion calibrated to the measured absorption at 300 K (see S7 and **Fig. S7** for details)[56,57]. This map serves as a qualitative indicator of how the material absorption evolves with temperature and wavelength. When the emission peak positions are overlaid, the high-temperature peaks fall in regions of comparatively high absorption and remain in the ASE regime, whereas the low-temperature peaks move into a low-absorption region where edge-guided lasing is observed. The visual separation between these regimes is consistent with the threshold behavior in **Fig. 4b** and shows that lasing occurs only when the cavity resonance enters a sufficiently low-loss portion of the material spectrum.

**Conclusion**

In summary, we have experimentally demonstrated radiative-channel-driven topological lasing in an active valley photonic crystal composed of isolated InP nanorods on an insulator, achieving stable room-temperature single-mode operation within a compact ~4λ-scale cavity. By explicitly embracing the non-Hermitian nature of the system—where gain, radiation leakage, and material absorption coexist—this work moves beyond conventional gain-centric descriptions and reveals loss channels as an active and decisive element in topological lasing. Through loss-included simulations combined with geometry- and temperature-dependent experiments, we show that lasing occurs only within a narrow spectral window where the above-light-line segment of the topological edge dispersion aligns with the gain–loss balance while remaining spectrally isolated from bulk modes. This radiative regime, rather than high-Q confinement, governs mode selection and stabilizes single-mode operation. Spatially resolved off-edge pumping further provides direct experimental evidence that the emission originates from edge-guided transport, excluding trivial photoluminescence or scattering-induced localization. Berry-curvature analysis confirms that the interface states arise from valley-Hall topology dictated by unit-cell symmetry breaking in the triangular lattice. More broadly, our results establish a radiative-channel framework for active topological photonics, in which loss is not merely a parasitic limitation but a design parameter that cooperates with topology to select and stabilize lasing pathways. The simple, substrate-supported nanorod architecture demonstrated here offers a practical and scalable route for implementing these concepts on chip, and provides a foundation for future active topological devices in which gain, radiation, and topology are co-engineered to control light generation and transport.

**Methods**

Sample fabrication: The VPC nanolaser devices were fabricated on a fully substrate-supported InP-on-insulator platform consisting of a 500-nm InP layer wafer-bonded onto a Si substrate covered by 2-µm $SiO_2$ layer for optical isolation towards the substrate[58], serving to confine the VPC modes in the higher index InP layer. To define the triangular-lattice nanorod patterns, we first deposited a ~3 nm $SiO_2$ adhesion layer by ALD and then spin-coated hydrogen silsesquioxane (HSQ) as a negative-tone electron-beam resist. The patterns were written using an electron-beam lithography system (Vistec EBPG 5200+) and developed to form a cross-linked HSQ mask. A short RIE step was used to remove the adhesion layer in the exposed areas, after which the HSQ mask was used directly for pattern transfer into InP. The nanorods were etched using an ICP-RIE tool (Oxford Instruments PlasmaLab 100)[59]. After etching, the sample was briefly dipped in diluted phosphoric acid ($H_3PO_4$:$H_2O$ = 1:10) to clean and passivate the etched surface, and then capped with a ~3 nm $Al_2O_3$ layer deposited by ALD.

Optical characterization: Devices were characterized by micro-photoluminescence (µ-PL) at room temperature under ambient conditions. Optical pumping was provided by a picosecond pulsed laser source (NKT Photonics SuperK Extreme) operated at a repetition rate of 78 MHz, with the pump wavelength set to 720 nm for above-bandgap excitation of InP. The pump beam was focused through a 50× objective (Mitutoyo M Plan Apo NIR, NA = 0.65) to a spot diameter of ~3 µm, and the emission was collected by the same objective. Incident pump power was measured at the sample plane, and power density was calculated using the spot area. The collected signal was filtered using a long-pass (or notch) filter to suppress residual pump light and analyzed using a grating spectrometer (Princeton Instruments Acton SpectraPro SP-2500) with an InGaAs line-array detector (Princeton Instruments PyLoN-IR). Real-space emission images were recorded on an IR camera/CCD in parallel with spectral acquisition. A localized pump spot is positioned approximately 1 µm away from the cavity edge, where it generates carriers and photoluminescence in the InP nanorods around the gain wavelength (~920 nm). This off-edge excitation creates a localized gain region that feeds the cavity eigenmodes.. The photonic crystal is designed to support guided topological edge modes at the emission wavelength, whereas it does not support guided modes at the pump wavelength (720 nm). As a result, off-edge pumping suppresses trivial edge-localized photoluminescence that can arise under global excitation, enabling a clear identification of edge-guided lasing. Low-temperature measurements (80–100 K) were performed in a high-vacuum chamber (base pressure ~$5 \times 10^{-6}$ Torr).

Loss-included electromagnetic simulations: To analyze mode selection and compare directly with experiments, we computed the $k_x$-resolved projected band structure using three-dimensional finite-difference time-domain (FDTD) simulations (Ansys Lumerical). Because our cavity is formed from a waveguide-based interface, the optical modes can be analyzed in terms of the dispersion along the propagation direction. Periodic (Bloch) boundary conditions were applied along the interface direction (x-axis), while perfectly matched layers (PML) were used along the transverse directions (y and z axes) to emulate open boundaries and absorb outgoing radiation. In the simulations, the InP material was assigned the experimentally measured complex refractive index $\tilde{n} = n + i\kappa$ (Fig. S4) to include both radiative leakage and material absorption. The projected bands were obtained by exciting the structure with a broadband source and extracting the spectral response as a function of $k_x$. Resonance frequencies and quality factors $Q(k_x)$ were extracted from the time-domain field response using Lumerical's built-in harmonic-inversion-based resonance analysis, which fits the recorded time trace to exponentially decaying oscillatory components to obtain the decay rate and corresponding $Q$. The edge branch was identified by its spectral separation from bulk bands in the projected band diagram.

**Data availability**

The data that support the findings of this study are available from the corresponding author upon reasonable request.

**Acknowledgements**

We thank the Cleanroom Operations Team of the Binnig and Rohrer Nanotechnology Center (BRNC) for their help and support, Daniele Caimi for performing the wafer bonding, and Dr Simone Iadanza for helpful discussions. This work was supported by the National Research Foundation of Korea (NRF) (Grant No. NRF-2020R1I1A3071811), the Hanbat National University Financial Accounting Research Fund (2024), the Swiss National Science Foundation (Grant Nos. 188173 and IZSEZ0_232489), and the EU ITN EID project CORAL (Grant Agreement No. 859841). Additional support was provided by the Basic Science Research Program through the National Research Foundation of Korea (NRF), funded by the Ministry of Education (Grant No. 2022R1A6A3A03069115).

**Author contributions**

S.K. conceived the idea, designed the study, performed the simulations, fabricated the devices, established the experimental setup, carried out the optical measurements, analyzed the data, and wrote the manuscript. M.S. and J.D. contributed to the experimental setup, fabrication support, and discussions. H.S. contributed to material growth and discussions. K.M. and C.-W.L. contributed to the interpretation of the results, revision of the manuscript, and supervision of the project. All authors reviewed and commented on the manuscript.

**Conflict of Interest**

The authors declare no conflict of interest.

**Supplementary Information**

# Radiative-channel valley topological laser

Seonyeong Kim[1,2,3], Markus Scherrer[4], Jakub Dranczewski[4,5], Heinz Schmid[4], Kirsten Moselund[2,3], and Chang-won Lee[1]

[1]Institute of Advanced Optics and Photonics, Hanbat National University, Daejeon, 34158, Korea

[2]Paul Scherrer Institute, Forschungsstrasse 111, Villigen 5232, Switzerland

[3]École Polytechnique Fédérale de Lausanne (EPFL), STI INPhO, Lausanne, Switzerland

[4]IBM Research Europe – Zürich, Säumerstrasse 4, Rüschlikon 8803, Switzerland

[5]Blackett Laboratory, Department of Physics, Imperial College London, SW7 2BW London, UK

## S1. Berry-curvature calculations for different unit-cell representations.

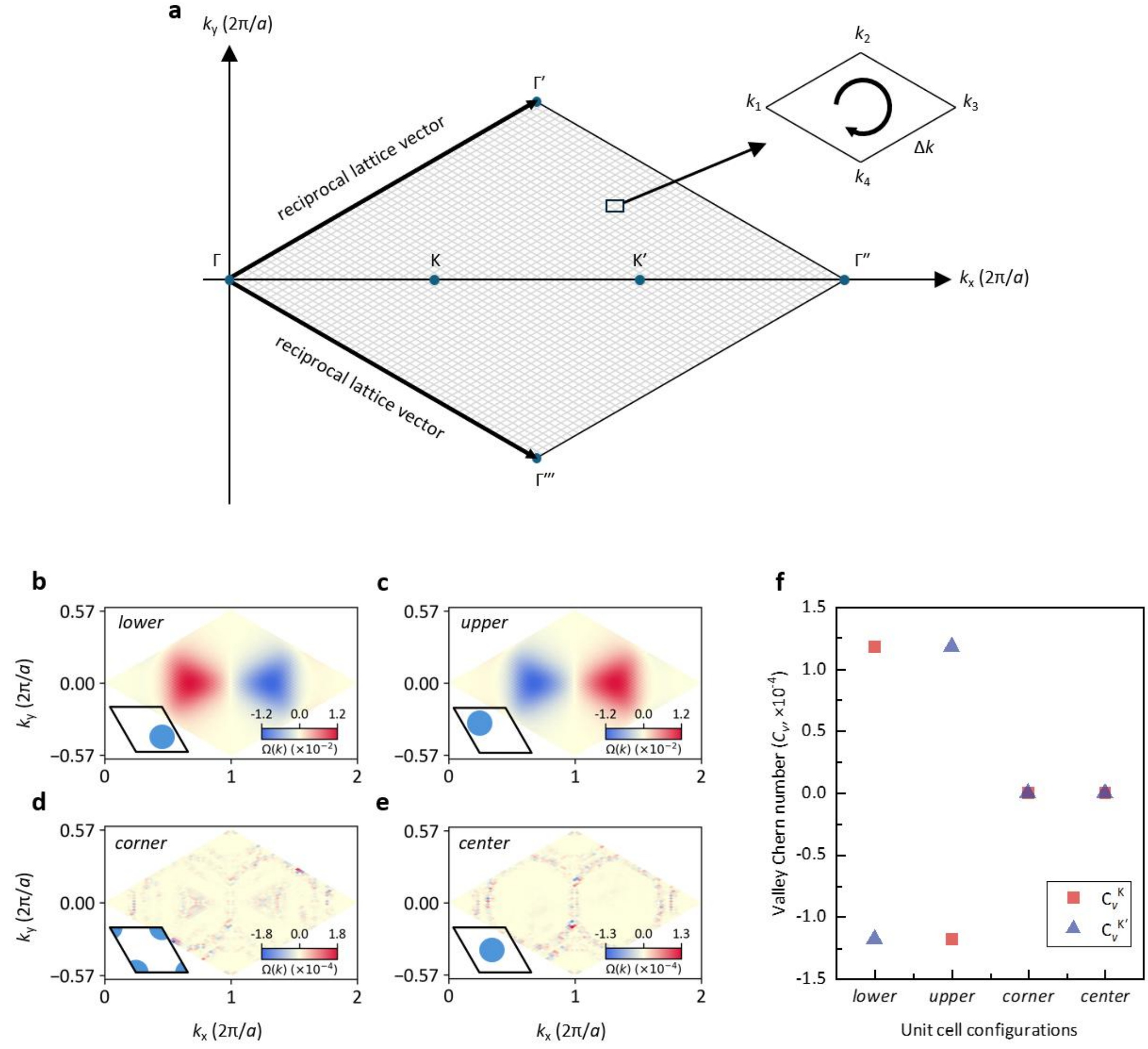

**Figure S1 | Berry-curvature distributions and valley Chern numbers for different unit-cell representations of a triangular photonic crystal.**

**a.** Rhombic representation of the first Brillouin zone defined by the reciprocal lattice vectors. The K and K′ points are indicated, and the Berry curvature is evaluated on a discretized k-space mesh using a plaquette-based formulation (inset). **b–e.** Berry-curvature distributions $\Omega(\mathbf{k})$ calculated for four different unit-cell representations: *lower*, *upper*, *corner*, and *center*. Insets show the corresponding unit-cell configurations used in the calculations. **f.** Valley Chern numbers $C_v^K$ and $C_v^{K'}$ obtained by integrating the Berry curvature over the corresponding valley regions for each unit-cell configuration.

We consider a two-dimensional photonic crystal (PhC) composed of dielectric rods in air, which has a photonic band gap under transverse-magnetic (TM) polarization. In this case, the magnetic field is confined to the in-plane components $(H_x, H_y)$, and the electric field becomes a scalar function; only one non-zero out-of-plane component $E_z$. We refer to $E_z$ simply as $E$ in the following discussion for brevity.

We first extract the electric field components $E_{\mathbf{k}}(\mathbf{r})$ for a unit cell at a given Bloch wavevector **k** via Finite-Difference Time-Domain (FDTD) method (Lumerical Inc.). Here, for simplicity, we assume a non-magnetic ($\mu = 1$), lossless, isotropic, and dispersionless medium. The Bloch boundary condition ensures that the resulting steady-state frequency-domain electric fields satisfy the Bloch condition for any lattice vector **R**:

$$E_{\mathbf{k}}(\mathbf{r}+\mathbf{R}) = e^{i\mathbf{k}\cdot\mathbf{R}} E_{\mathbf{k}}(\mathbf{r}) \quad (1)$$

This relation implies that $E_{\mathbf{k}}(\mathbf{r})$ takes the form of a Bloch-mode solution,

$$E_{\mathbf{k}}(\mathbf{r}) = e^{i\mathbf{k}\cdot\mathbf{r}} u_{\mathbf{k}}(\mathbf{r}) \quad (2)$$

where $u_{\mathbf{k}}(\mathbf{r})$ is a lattice periodic function satisfying $u_{\mathbf{k}}(\mathbf{r}+\mathbf{R}) = u_{\mathbf{k}}(\mathbf{r})$.

**Berry curvature calculation**

To evaluate the Berry curvature $\Omega(\mathbf{k})$, we discretize the first Brillouin zone (BZ) into a uniform 52 × 52 mesh in reciprocal space and obtain the $E_{\mathbf{k}}(\mathbf{r})$ over the entire BZ. For convenience in subsequent calculations, we adopt a rhombus-shaped representation of the BZ defined by the reciprocal lattice vectors, as illustrated in **Figure S1a**[1,2].

In a discretized context, the Berry curvature over each plaquette is approximated as the Berry phase $\phi$ accumulated around a small plaquette in $k$-space divided by its area, that is $\Omega(\mathbf{k}) \approx \phi/\Delta k_x \Delta k_y$, where $\Delta k_x$ and $\Delta k_y$ denote uniform grid spacing along the two sampling directions. We focus on the lowest band where no degeneracy is present. The Berry phase $\phi$ associated with a plaquette defined by four neighboring $k$-points $\{k_1, k_2, k_3, k_4\}$ is computed using four-point formula:

$$\phi = -\mathrm{Im}\ln\left[\langle u_{k_1}(\mathbf{r})|u_{k_2}(\mathbf{r})\rangle\langle u_{k_2}(\mathbf{r})|u_{k_3}(\mathbf{r})\rangle\langle u_{k_3}(\mathbf{r})|u_{k_4}(\mathbf{r})\rangle\langle u_{k_4}(\mathbf{r})|u_{k_1}(\mathbf{r})\rangle\right] \quad (3)$$

Here, $u_{k_i}(\mathbf{r})$ denotes the normalized periodic part of the Bloch mode at each plaquette corner, obtained by removing the Bloch phase factor $e^{i\mathbf{k}\cdot\mathbf{r}}$ from Eq. (2). The normalization convention is used in the following form:

$$\langle u_{\mathbf{k}}(\mathbf{r})|u_{\mathbf{k}'}(\mathbf{r})\rangle = \sum_{i,j} [\varepsilon(i,j) u_{\mathbf{k}}(i,j)]^* u_{\mathbf{k}'}(i,j)\Delta S \quad (4)$$

where $(i,j)$ indexes the discretized real-spatial grid points in the unit cell, $\varepsilon(i,j)$ is the local dielectric constant, and $\Delta S$ is the area of a real-space grid cell. The normalization condition $\langle u_{\mathbf{k}}|u_{\mathbf{k}}\rangle = 1$ is enforced for all **k**.

Using the above formalism, we evaluate the Berry curvature for several unit-cell representations of a triangular lattice photonic crystal, as illustrated by the insets in **Fig. S1b–e**. Although these representations describe the same infinite lattice when periodically repeated, the placement of the photonic atom within the unit cell determines whether inversion symmetry is broken and, consequently, how the Berry curvature is distributed in momentum space.

**Fig. S1b** and **S1c** show the Berry-curvature distributions for two inversion-related unit-cell configurations, referred to as the *lower* and *upper* configurations, in which inversion symmetry is explicitly broken by placing a single photonic atom at inequivalent positions within the rhombic unit cell. In both cases, the Berry curvature acquires a finite magnitude and is strongly localized around the K and K′ valleys with opposite signs, indicating a well-defined valley-contrasting topology. By contrast, **Figs. S1d** and **S1e** present Berry-curvature maps for alternative unit-cell representations (*corner* and *center* configurations) of the same triangular lattice, for which inversion symmetry is effectively preserved by symmetry-equivalent placement of the photonic atom. In these cases, the Berry curvature is no longer concentrated at the valley points but instead spreads over a broader region of the Brillouin zone, leading to the absence of a well-defined valley-localized Berry curvature.

**Valley Chern number evaluation**

To provide a quantitative measure of the valley-dependent topology inferred from the Berry-curvature distributions, we evaluate the valley Chern number. The BZ is divided into two valley regions $\mathcal{V}_{\mathrm{K}}$ and $\mathcal{V}_{\mathrm{K}'}$ by the $\Gamma'$–$\Gamma'''$ boundary (**Fig. S1a**), yielding the corresponding valley Chren numbers $C_v^K\,and\,C_v^{K'}$ defined as

$$C_v^{K(K')} = \frac{1}{2\pi} \sum_{\mathbf{k}_{ij} \in \mathcal{V}_{K(K')}} \Omega(\mathbf{k}_{ij})\, \Delta k_x \Delta k_y.$$

where $\Omega(\mathbf{k}_{ij})$denotes the Berry curvature assigned to each plaquette, and $\Delta k_x \Delta k_y$is the plaquette area.

**Fig. S1f** summarizes the valley Chern numbers obtained for different unit-cell representations of the triangular PhC. Owing to time-reversal symmetry, the total Chern number $C = C_v^K + C_v^{K'}$ remains zero in all cases, indicating the absence of a nonzero global topological invariant. Nevertheless, when inversion symmetry is broken, the Berry curvature becomes redistributed in momentum space, allowing for a nontrivial valley Chern number, which serves as a valley topological invariant[3].

For the lower and upper unit-cell configurations, the valley Chern numbers are finite and have opposite signs, even though the total Chern number remains zero. According to the principle of bulk–edge correspondence, a difference in topological invariants between two adjacent bulk domains necessitates the existence of localized edge modes at their interface[4]. A finite contrast in valley Chern numbers between inversion-related unit cells provides the topological condition for the emergence of valley-Hall edge states at domain walls [3]. Conversely, when the valley Chern number vanishes for *corner* and *center* configurations, such topological edge states are not expected from the valley-Hall mechanism. This result demonstrates that a triangular lattice photonic crystal can be regarded as a valley photonic crystal when an appropriate inversion-breaking unit cell is employed, highlighting unit-cell selection as a key design principle for realizing and engineering valley-Hall topological edge states in triangular-lattice photonic systems.

**S2: Dependence of edge-state dispersion and localization on unit-cell configuration and domain alignment.**

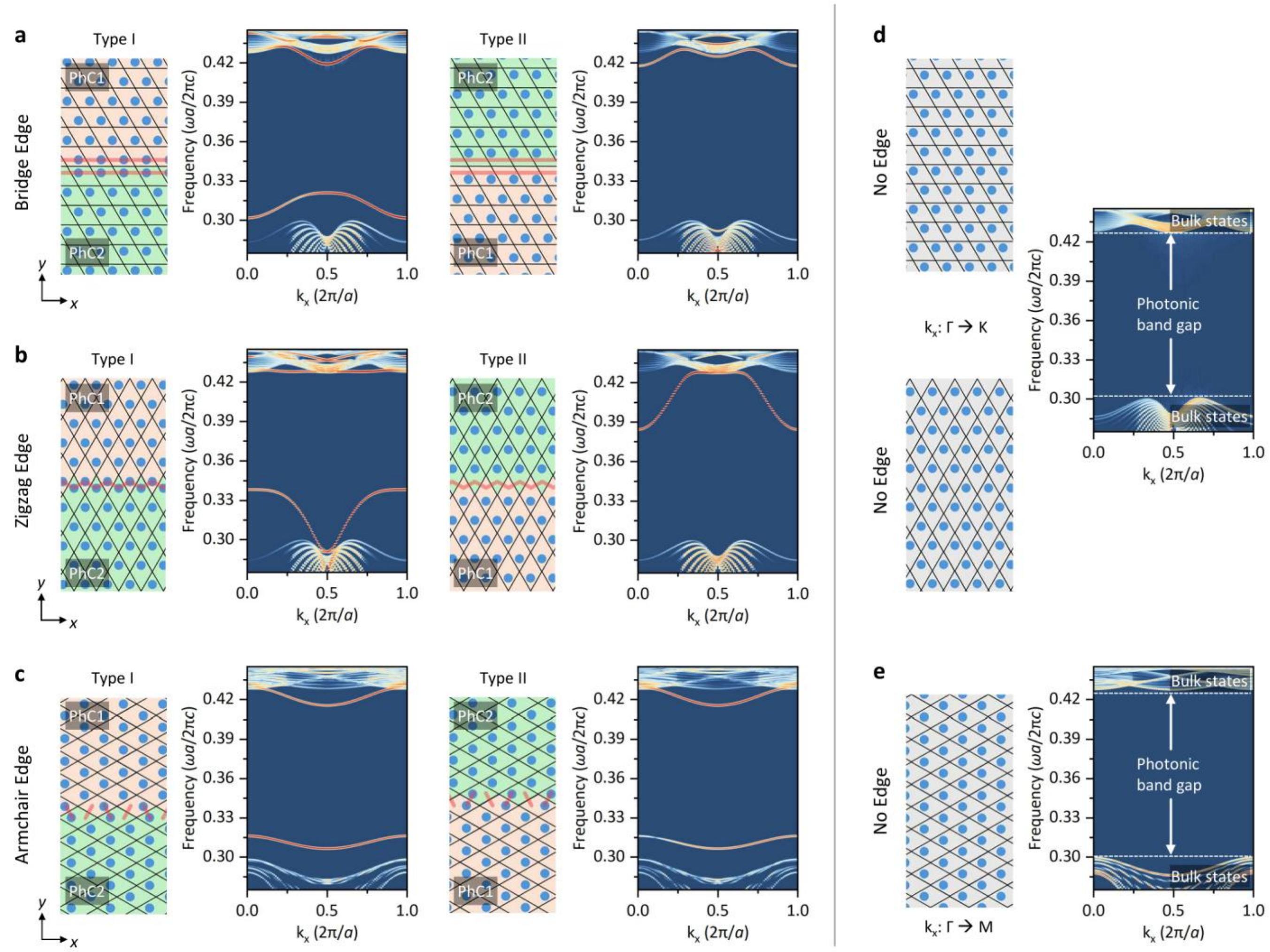

**Figure S2 | Edge geometries and projected band structures in a triangular-lattice photonic crystal.**

**a–c**, Schematics of the bridge (BE), zigzag (ZE), and armchair (AE) edge geometries, together with their corresponding projected band structures for the two interface types (Type 1 and Type 2). **d,e**, Projected band structures of the edge-free supercells, shown as bulk references. The blue–red color scale indicates the spectral intensity, with red corresponding to higher intensity. A group of dipole sources is placed adjacent to the interface to preferentially excite edge-localized modes, which therefore appear with higher intensity than bulk modes.

**Figure S2** summarizes how topological edge dispersions in a triangular-lattice photonic crystal depend on the edge configurations. The figure illustrates three representative edge types—bridge (BE, **Fig. S2a**), zigzag (ZE, **Fig. S2b**), and armchair (AE, **Fig. S2c**)—together with their corresponding projected band structures calculated using FDTD simulations. For each edge geometry, two interface types (Type 1 and Type 2) are defined by exchanging the upper (PhC1) and lower (PhC2) unit-cell configurations across the interface, while keeping all bulk parameters identical. To focus on the existence, shape, and spectral position of the edge dispersions over the frequency range of interest, material loss was set to be negligibly small, avoiding band blurring that would otherwise obscure individual photonic states. The simulations were performed in two dimensions to exclude radiation leakage into the substrate and air, while also reducing the computational cost, without affecting the qualitative features of the edge dispersions.

Comparison with the projected band structures of the edge-free supercells shown in **Figs. S2d,e** is useful for identifying edge-localized modes. In the absence of an interface, only bulk states appear, clearly defining the lower and upper band edges as well as the photonic band gap, which provides a clear reference for analyzing the edge-state dispersions in **Figs. S2a–c**. Notably, the BE and ZE share the same projected band structure, whereas the AE exhibits a distinct projection. This difference originates from the supercell orientation: the BE and ZE geometries correspond to a Γ–K projection, while the AE geometry corresponds to a Γ–M projection. Against this bulk reference, edge-localized states are clearly visible within the band gap, appearing as high-intensity features highlighted by the red-colored contours in the **Figs. a–c**. While the BE and ZE edges exhibit markedly different edge dispersions between Type 1 and Type 2, the AE shows nearly identical dispersions for the two types, consistent with the fact that the two AE configurations are related by a simple flip with respect to the vertical y-axis. Since the position and shape of the edge dispersion determine fundamental characteristics of guided modes, such as their spectral range and group velocity, careful consideration of the edge dispersion is essential in the design of waveguide-based photonic devices. In the following device studies, the bridge edge (BE) is used as a representative edge geometry.

**S3: Band-edge laser characteristics**

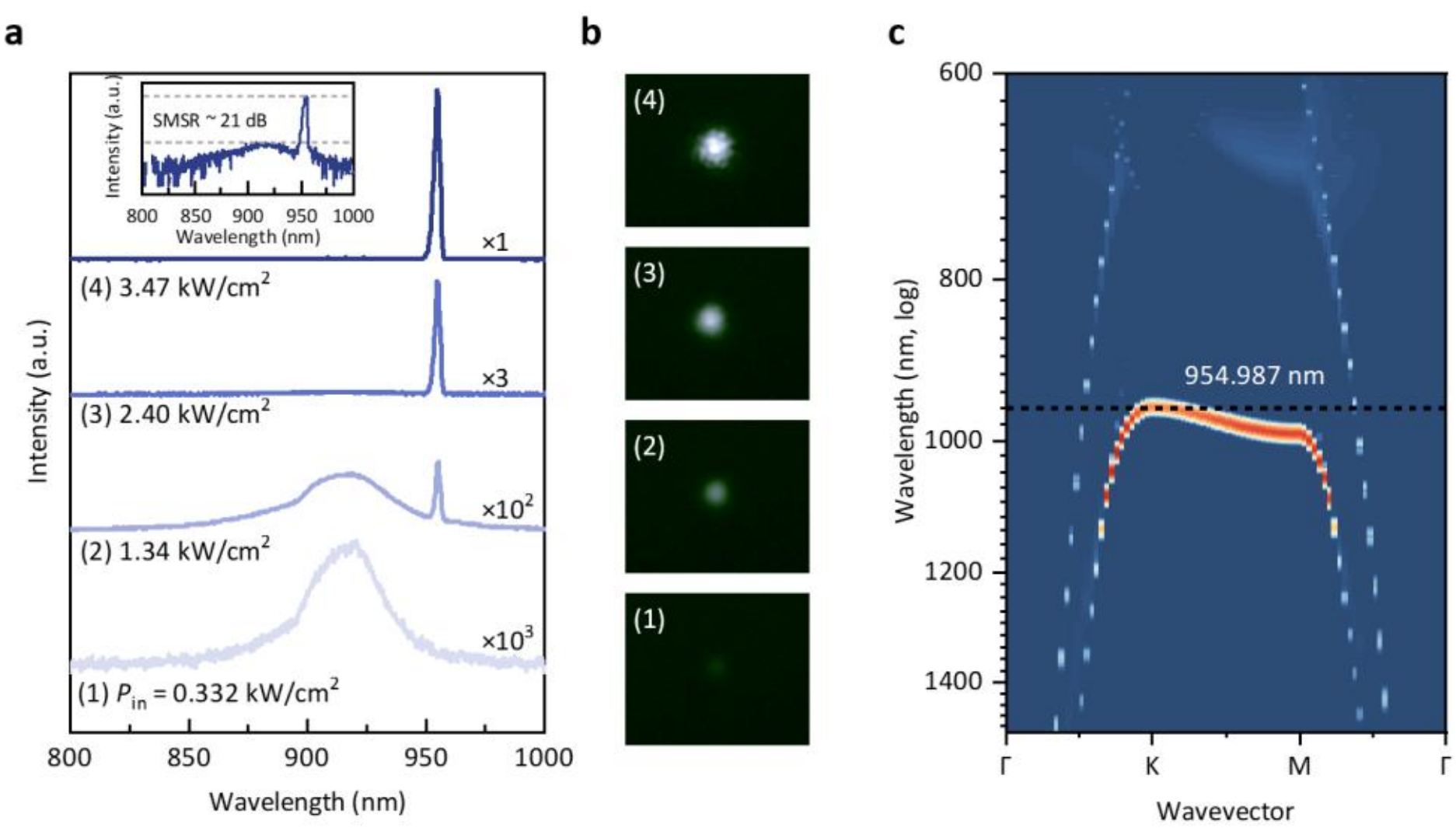

**Figure S3 | Power-dependent emission characteristics of the band-edge laser.**

**a**. Emission spectra measured from the uniform photonic-crystal region under increasing pump power densities, showing the transition from broadband spontaneous emission to narrowband lasing above threshold. **b**. Corresponding real-space emission images acquired at the same pump conditions as in (a). In contrast to the topological edge laser (Fig. 2b), the above-threshold emission remains spatially confined to the excitation spot. **c**. Calculated photonic band structure of the uniform triangular-lattice photonic crystal. The dashed line indicates the lasing peak, which lies near the band edge of the bulk photonic crystal.

**Table S1 | Performance comparison of reported valley topological lasers.**

| Ref. | Wavelength | Threshold | Single Mode | Device Footprint | Substrate-supported | Pumping | |
|---|---|---|---|---|---|---|---|
| | | | | | | Type | Geometry |
| Zeng et al.[5] 2020 | ~ 94 μm (~ 3.2 THz) | Current-driven | No | ~ 4λ | Yes | Electrical | Edge-shaped |
| Noh et al.[6] 2020 | ~ 1550 nm | ~ 0.17 kW/cm$^2$ | Yes | ~ 14λ | No | Optical | Global |
| Smirnova et al.[7] 2020 | ~ 1400 nm | 66 μJ/cm$^2$ ~ 8.3 kW/cm$^2$ | Yes | ~ 2.3λ | No | Optical | Global |
| Peng et al.[8] 2024 | ~ 430 nm | 15.2 μJ/cm$^2$ ~ 6.08x10$^4$ kW/cm$^2$ | Yes | ~ 30 λ | No | Optical | Edge-shaped |
| Hong et al.[9] 2025 | 1610 nm | 29 kW/cm$^2$ | Yes | ~ 4λ | No | Optical | Global |
| **This work** | **~ 950 nm** | **2.4 kW/cm$^2$** | **Yes** | **~ 4λ** | **Yes** | **Optical** | **Off-edge local spot** |

The device footprint is expressed in units of the emission wavelength λ. "Substrate-supported" denotes membrane-free device architectures that do not rely on suspended photonic crystal slabs. Threshold values are quoted as reported in the literature and may correspond to different pumping schemes.

**S4 : Optical constants of the InP film**

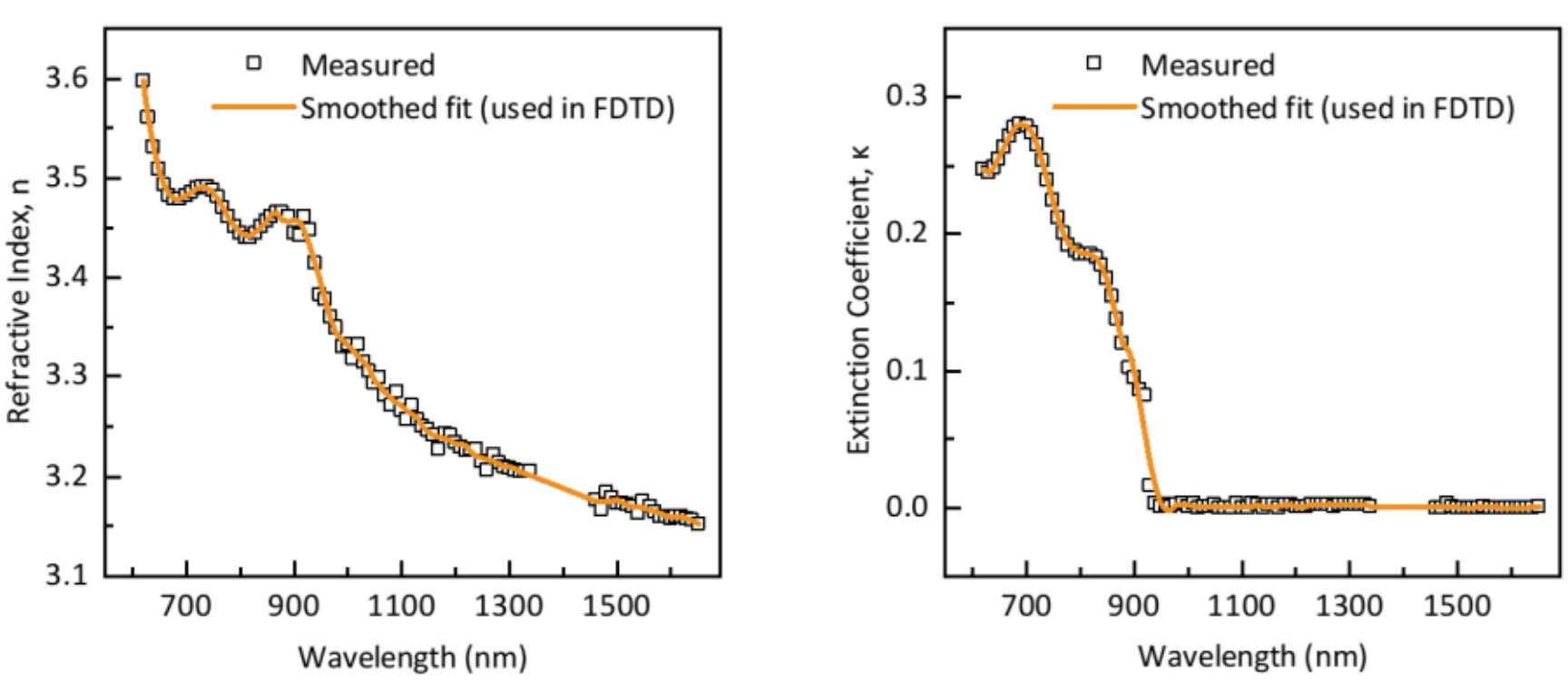

**Figure S4 | Measured complex refractive index of the InP film prior to patterning.**

The real (n) and imaginary (κ) parts of the refractive index were measured by spectroscopic ellipsometry on an unpatterned 500-nm-thick InP film on $SiO_2$/Si at room temperature under ambient conditions. Symbols denote the experimentally measured data. Solid lines represent smoothed and interpolated fits to the measured data, which were directly used as the material input for the FDTD simulations.

## S5: Effect of material absorption on the topological edge mode

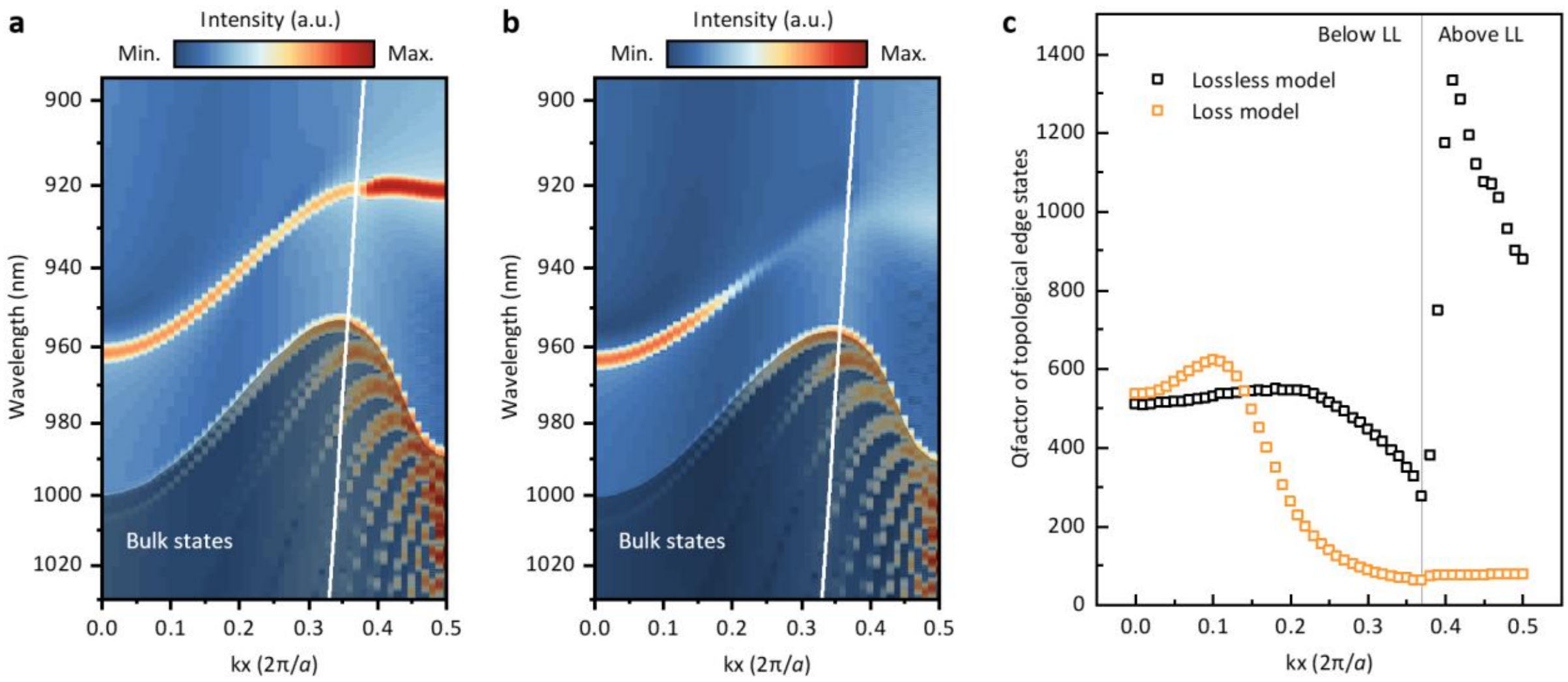

**Figure S5 | Effect of material absorption on the projected band structure and Q factor of the topological edge mode.**

**a,** Projected band structure of the topological interface supercell calculated using the experimentally measured refractive index (Fig. S4), with the material loss artificially set to zero ($\kappa = 0$) for comparison. **b,** Same calculation including the measured material loss ($\kappa(\lambda)$). The white line indicates the light line (LL). **c,** Q factor of the edge mode extracted from (a) and (b). While the bulk bands are nearly identical and the edge-mode dispersion retains a similar shape and spectral position, the low-wavelength region shows a clear contrast: in the lossless case the edge branch appears sharper and exhibits a higher Q below the light line, whereas including material absorption reduces the Q factor and renders the edge mode less distinct.

## S6: Temperature dependence of the reference InP photoluminescence.

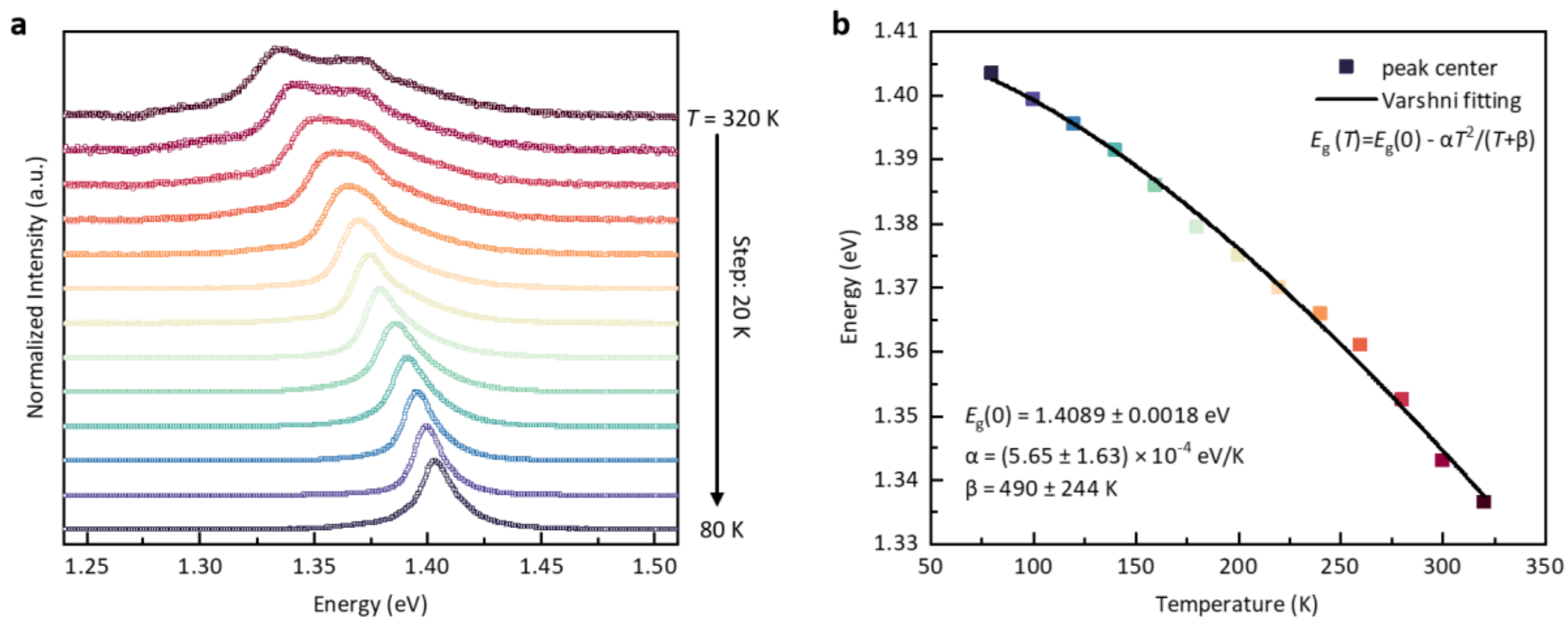

**Figure S6 | Temperature-dependent PL spectra and Varshni fit of InP.**

**a**, Normalized PL spectra measured from an unpatterned InP region between 80 and 320 K, showing a monotonic redshift of the band-to-band emission with increasing temperature. Spectra are vertically offset for clarity. **b**, Extracted peak energies as a function of temperature (symbols) and a fit using the Varshni relation (solid line)[10].

## S7: vRS-based construction of a temperature-dependent absorption surrogate

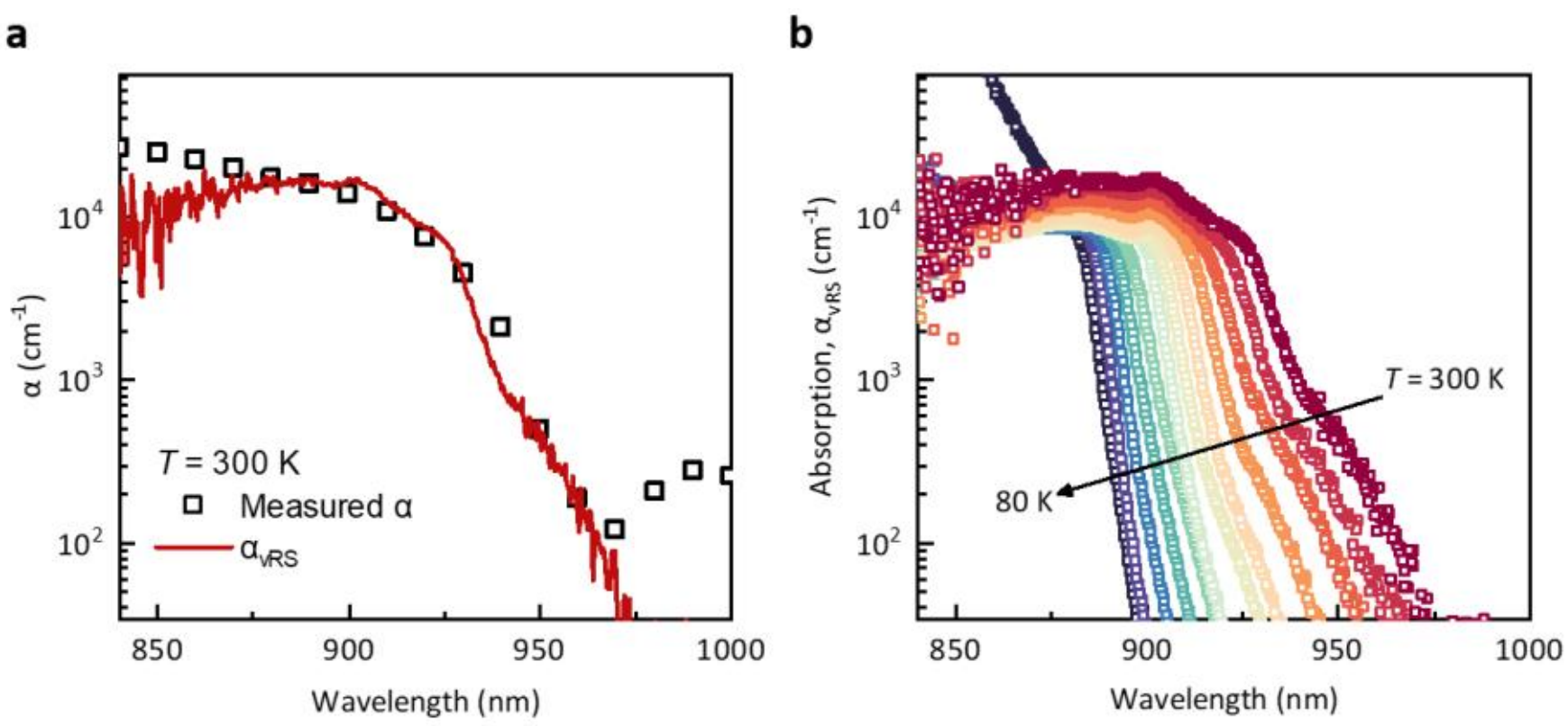

**Figure S7 | Construction and calibration of the vRS-derived absorption.**

**a,** Comparison between the independently measured absorption coefficient at 300 K (black squares) and the absorption surrogate obtained from flat-region photoluminescence via the vRS-based conversion (red line). The two spectra are matched near the band edge using a single global scale factor. **b,** Temperature series of the calibrated vRS-derived absorption spectra from 80 to 300 K (20 K steps), illustrating the systematic temperature evolution of the inferred absorption profile.

To track the temperature evolution of material loss, we convert the flat-region photoluminescence (PL) into a van Roosbroeck–Shockley (vRS)–based surrogate of absorption according to[11,12]

$$\alpha_{\mathrm{vRS}}(E,T) = C\,\frac{I_{\mathrm{PL}}(E,T)\ \exp(E/k_{\mathrm{B}}T)}{n^2(E)\,E^2}$$

, where $I_{\mathrm{PL}}(E,T)$ is the measured PL intensity, $n(E)$ is the refractive index, and $C$ is a global scale factor determined experimentally. Non-positive values arising from noise are discarded. This expression is based on the vRS detailed-balance relation evaluated for the flat-region PL, assuming quasi-thermal carrier distributions and optical prefactors that vary only slowly with energy within the limited spectral range of interest. The conversion is therefore used strictly as a surrogate to capture the absorption lineshape and its relative temperature dependence. To suppress multiplicative noise while preserving the band-edge flank, the spectra are analyzed on a logarithmic scale, $y = \log_{10} \alpha_{\mathrm{vRS}}$, and mild local adjacent-averaging smoothing is applied only near the rising edge. The operational band edge $E_g(T)$ is defined as the energy at which $dy/dE$ is maximized, corresponding to the most well-defined and reproducible feature of the absorption onset. All spectra are energy-aligned to $E_g(T)$, which provides a physically motivated normalization that is least sensitive to background offsets and long-wavelength tail variations.

As shown in **Fig. S8a**, the scale factor $C$ is obtained at 300 K by calibrating $\alpha_{\mathrm{vRS}}$ to the independently measured absorption coefficient from ellipsometry (**Fig. S4**) within a near-edge region of interest, where the PL–absorption correspondence is most reliable, avoiding deep sub-gap and high-energy regions where noise and reabsorption effects become significant. The same value of $C$ is then fixed and applied to all temperatures, yielding a calibrated family $\alpha_{\mathrm{vRS}}^{\mathrm{cal}}(E,T)$, as shown in **Fig. S8b**. This procedure anchors the surrogate to experiment while preserving strictly relative temperature evolution. Although the resulting $\alpha_{\mathrm{vRS}}$ does not represent an absolute absorption coefficient, this procedure provides a consistent basis for tracking the relative evolution of the absorption spectrum with temperature.